# Tracing out the Berry curvature dipole and multipoles in second harmonic Hall responses of time-reversal symmetric insulators

*Mahmut Sait Okyay[1], Shunsuke Sato[2,3], Kunwoo Kim[4], Binghai Yan[5], Hosub Jin[1], Noejung Park[1*]*

[1]*Department of Physics, Ulsan National Institute of Science and Technology (UNIST), Ulsan, 44919 Republic of Korea*

[2]*Center for Computational Sciences, University of Tsukuba, Tsukuba 305-8577, Japan*

[3]*Max Planck Institute for the Structure and Dynamics of Matter, 22761 Hamburg, Germany*

[4]*Department of Physics, Chung-Ang University, 06974 Seoul, Korea*

[5]*Department of Condensed Matter Physics, Weizmann Institute of Science, Rehovot 7610001, Israel*

[*]*Correspondence and requests for materials should be addressed to (email: noejung@unist.ac.kr)*

## ABSTRACT

Various nonlinear characteristics of solid states, such as the circular photogalvanic effect of time-reversal symmetric insulators, the quantized photogalvanic effect of Weyl semimetals, and the nonlinear Hall effect of time-reversal symmetric metals, have been associated with the Berry curvature dipole (BCD). Here, we explore the question of whether the Berry curvature dipole and multipoles of time-reversal symmetric insulators can be traced in the nonlinear optical responses. We performed real-time time-dependent density functional theory calculations and examined the second harmonic generation susceptibility tensors. The two-band term of the susceptibility tensor is sharply proportional to the interband BCD, dominating over the Hall response once the cancellation effect of the multiple reflection symmetries is lifted. We suggest that the nonlinear Hall component of the second-harmonic spectra of insulators can also be utilized as an effective tool to extract the band structure geometry through Berry curvature dipole and possibly multipoles.

## I. INTRODUCTION

Many key discussions of recent studies of condensed matter physics, having gone beyond the energy-momentum relations of the eigenvalues, have mainly focused on the geometrical nature of quantum mechanical wavefunctions of electronic band states. Among the prime attributes underlying such geometrical natures is the Berry curvature of the Bloch state [1]. A transcendental recognition of the importance of the Berry curvature was earlier provided by Thouless, Kohmoto, Nightingale, and Nijs (TKNN), who proved that the transverse charge conductivity of a two-dimensional (2D) band insulator is quantized in terms of the zone-integrated Berry curvature [2]. In parallel with the developing geometrical notions of the Landau level states in integer quantum Hall effects [3,4], TKNN suggested that non-zero integer multiples of the "anomalous Hall conductivity" is allowed only when the time-reversal symmetry (TRS) is broken [2,5]. Similar topological notions have later developed even for the time-reversal symmetric solid states in the context of the quantum spin Hall states [6-8] and the Weyl/Dirac semimetals, which is reminiscent of the relativistic massless particle with distinct helicities [9,10].

For a time-reversal symmetric system, the charge Hall current vanishes in the linear regime and the net Berry curvature is always integrated out because of the odd symmetry in the momentum space. Even under such a constraint of TRS, the effect of the Berry curvature is still appreciable in nonlinear optical responses. As suggested by Sodemann and Fu in 2015 [11] and as measured by Ma et al. in 2019 [12], an inversion-broken metallic solid exhibits a nonzero charge Hall effect in the second-order response, which can be associated with the Berry curvature dipole (BCD) – the Berry curvature integrated over the metallic Fermi surface. On the other hand, for insulators, to describe a circular photogalvanic effect in response to a circularly polarized photoexcitation [13-17], the concept of the interband BCD has been devised by the integration of the Berry curvature difference of two band-edge states over the surface with a given energy separation. It is further noteworthy that the circular photogalvanic effect is quantized at nodes of Weyl semimetals in terms of interband BCD [18]. The concept of the BCD has also been extended to discussions for the magnetoresistance of Weyl metals [19] and the orbital Edelstein effect [20].

Interband photogalvanic effects are allowed only when the light frequency is larger than the bandgap in time-reversal symmetric conditions [13,14,17], and thus the energy surface for the integration of the interband BCD has not yet been defined with the sub-bandgap driving field in the literature. In contrast with these photocurrents, the second harmonic generation (SHG) is allowed even if the light frequency is below the bandgap. Here, we are motivated by the questions of whether time-reversal symmetric 2D insulators can produce a Hall response for sub-bandgap light frequency in analogy with the nonlinear Hall effect of metals, and whether the detected SHG spectra can be used to extract the intrinsic Berry curvature distributions by a proper definition of an interband BCD. We examine the nonlinear Hall responses of noncentrosymmetric 2D insulators to linearly polarized sub-bandgap frequency light fields in the framework of the real-time time-dependent density functional theory (rt-TDDFT). We show that the SHG susceptibility [13,21] can be decomposed into two-band and three-band terms. The former is proportional to the interband BCD, vanishing in the existence of the multiple reflection symmetries, while the latter is less affected by the material's spatial symmetry. Moreover, the SHG signals are governed by the interband BCD of band-edge states when the material has single mirror reflection. Once the system is driven by a strong field, such that high harmonic generation (HHG) can be observed non-perturbatively, the addition of higher-order contributions including Berry curvature multipoles to the BCD is required to reflect the correct SHG spectrum. Our results indicate that irrespective of metal or insulator, the nonlinear Hall responses not only indicate the broken-inversion symmetry but also contain information of the Berry curvature distribution.

## II. RESULTS AND DISCUSSION

### A. Second-order optical responses of insulators and nonlinear Hall effect of metals

Various optical responses of materials have usually been discussed in terms of dipole-approximated Hamiltonian: $\hat{H}(\hat{\mathbf{p}},t) = \hat{H}(\hat{\mathbf{p}} + \frac{e}{c}\mathbf{A}(t)) \approx \hat{H}(\hat{\mathbf{p}}) + \frac{e}{mc}\hat{\mathbf{p}}\cdot\mathbf{A}(t)$, where the time-dependent vector potential $\mathbf{A}(t) = -c\int dt\mathbf{E}(t)$ originates from the applied electric field $\mathbf{E}(t)$, and $\hat{\mathbf{p}}$ represents the canonical momentum operator [13]. For an adiabatically evolving Bloch state, the momentum operator can be replaced with the velocity operator $\hat{\mathbf{p}}/m = \partial\hat{H}(\mathbf{k})/\hbar\partial\mathbf{k}$, where $\hat{H}(\mathbf{k})$ is the corresponding k-resolved Hamiltonian. For insulators, the second order effect of the dipole term produces plentiful optical responses, including injection current, shift current, second harmonic generation, optical rectification, and electro-optic effect [13,14]. The shift current contains a net DC current on the absorption of a photon, which possesses substantial implications for photovoltaic applications. It also refers to as linear photogalvanic effect (LPGE) in time-reversal symmetric conditions [17]. The injection current has been discussed in the framework of the circular photogalvanic effect (CPGE), which signifies the generation of DC current rate bias in response to the absorption of a circularly polarized photon. The material symmetry has prominent role in determination of the current direction in both shift and injection currents. On the other hand, for metallic systems, the second-order effect has recently attracted substantial attention in the perspective of the nonlinear Hall effect (NLHE, also known as intraband LPGE), which is not constrained by the time-reversal symmetry in contrast to the topological quantum Hall states [2,11,12,17].

It is not yet discussed whether the analogous Hall response can be extended to cases of insulators. To proceed to a consistent discussion on this second-order effect, we need to specify the notations for the Berry curvature of insulators and metals. Given the energy band structure $\varepsilon_n(\mathbf{k})$ and the eigenstates $|\psi_n(\mathbf{k})\rangle$ of the static Hamiltonian $\hat{H}(\mathbf{k})$ for the Bloch state with the momentum vector $\mathbf{k}$, the Berry curvature of the band $n$ is defined as

$$\Omega_n(\mathbf{k}) = 2\,\mathrm{Im}\sum_{m\neq n}\frac{v^y_{nm}v^x_{mn}}{\omega^2_{mn}}, \qquad (1)$$

where $v^\alpha_{nm} = \langle\psi_n(\mathbf{k})|\partial\hat{H}(\mathbf{k})/\hbar\partial k_\alpha|\psi_m(\mathbf{k})\rangle$ represent the velocity matrix elements between bands $n$ and $m$ and $\omega_{mn} = [\varepsilon_m(\mathbf{k}) - \varepsilon_n(\mathbf{k})]/\hbar$. On the other hand, the interband Berry curvature between the bands $n$ and $m$ can be written as

$$\Omega_{nm}(\mathbf{k}) = 2\,\mathrm{Im}\,\frac{f_{n\mathbf{k}}(1-f_{m\mathbf{k}})v^y_{nm}v^x_{mn}}{\omega^2_{mn}}, \qquad (2)$$

where $f_{n\mathbf{k}}$ indicates the occupation factor, and we force the first (second) index $n$ ($m$) to be an occupied (unoccupied) state. This form of the Berry curvature has been defined for the interband transitions of wide-gap insulators in the framework of the CPGE [15,16].

**B. Second harmonic Hall effect of two-dimensional insulators**

We now challenge our first main question whether a time-reversal symmetric 2D insulator can produce Hall responses, which can be considered as an extension of the NLHE to insulators. We pursue the solution, as depicted in Fig. 1a, by investigating the transverse second-harmonic current $\mathbf{J}_\perp(t)$ (in $y$-direction) in response to a given driving field $\mathbf{E}(t) = E_0\,\mathrm{Re}[e^{i\omega t}]\hat{\mathbf{x}}$ oscillating in the *x*-direction with the frequency below the bandgap: $\hbar\omega < \varepsilon_{\text{gap}}$. Here, we assume the field strength $E_0$ is real for simplicity. Since the time-reversal symmetric insulators are not allowed to show net second-order DC response to a sub-bandgap light, we confine our study to the second-harmonic generation responses [13,14,22]. As studied in second-order optical effects, the transverse electric polarization can be written as $P_y(t) = -\chi^{(2)} E_0^2 \text{Im}[e^{i2\omega t}]$ which leads to the current response, $J_y(t) = \dot{P}_y(t) = -2\omega\chi^{(2)} E_0^2 \text{Re}[e^{i2\omega t}]$, where $\chi^{(2)} \equiv \text{Im}[\chi^{yxx}(-2\omega;\omega,\omega)]$ is the imaginary part of the SHG susceptibility [13]. Hereafter, this frequency-doubled transverse response is called as second harmonic Hall effect (SHHE).

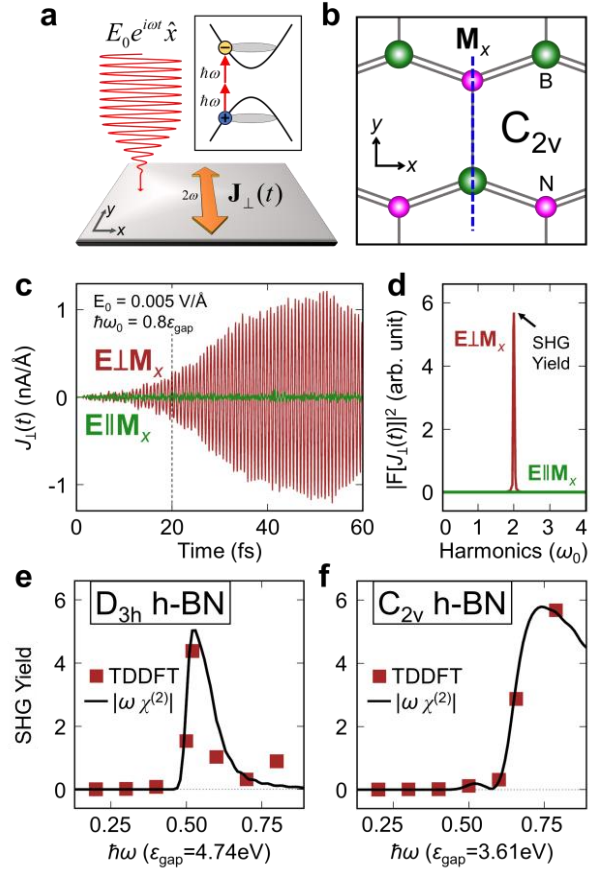

FIG. 1. The second-harmonic Hall current spectra in h-BN by rt-TDDFT calculations. **a**, The illustration of second harmonic Hall effect. The red line represents the applied liner light field polarized in the x-direction and the orange arrow shows the oscillating transverse SHHE current response. The inset shows the electron (blue ball)

and hole (yellow ball) carriers excited by two-photon absorption. **b** is the top view of the strained h-BN monolayer sheet. The blue dashed line represents the mirror plane $M_x$. **c**, The brown (green) oscillating line is the real-time profile of the transverse current density response obtained by rt-TDDFT calculations when the applied electric field is perpendicular (parallel) to the mirror plane. The black dashed line represents the time that the field is fully turned on, $t_0$ = 20 fs. **d**, The Fourier transformation spectra of the current densities given in **c** for the time interval of [20 fs,60 fs]. The x-axis is scaled with the applied field frequency, $\hbar\omega_0 = 0.8\varepsilon_{\text{gap}}$. The SHG yield is depicted with an arrow. **e** and **f** are the comparison of the rt-TDDFT yields and the theoretical SHG susceptibility spectra for three-fold symmetric pristine h-BN and two-fold symmetric strained h-BN, respectively. The light frequency is scaled with the bandgap of each structure. The brown filled squares are the SHG yields of the transverse current as given in **d**. The black line is the SHG susceptibility spectra multiplied by the light frequency, which is in same unit with the current, and the same scaling factor is used to fit SHG spectra to TDDFT data in **e** and **f**.

To realistically examine the oscillating current responses of the insulating materials, here we perform rt-TDDFT calculations. We apply a spatially uniform linearly polarized oscillating electric field, which can be viewed as a dipole-approximated continuous laser beam in the long-wavelength limit [23]:

$$\mathbf{E}(t) = f(t)E_0 \cos(\omega t)\hat{\mathbf{x}} \text{ or } f(t)E_0 \cos(\omega t)\hat{\mathbf{y}}, \tag{3}$$

where, $f(t)$ is the linear turning-on function, $f(t) = t/t_0$ when $t < t_0$ and $f(t) = 1$ when $t \geq t_0$, with $t_0$ = 20 fs. The time-varying k-resolved Hamiltonian can be determined with Peierls substitution $\hat{\mathbf{p}} \rightarrow \hat{\mathbf{p}} + \frac{e}{c}\mathbf{A}(t)$. Under the time-dependent driving field, the time-evolving Kohn–Sham (KS) Bloch states $|\psi_{n\mathbf{k}}(\mathbf{r},t)\rangle$ follow the time-dependent Kohn-Sham equation:

$$i\hbar\frac{\partial}{\partial t}|\psi_{n\mathbf{k}}(\mathbf{r},t)\rangle = \left[\frac{1}{2m_e}\left(\hat{\mathbf{p}} + \frac{e}{c}\mathbf{A}(t)\right)^2 + \hat{V}_{\text{atom}} + \hat{V}_{\text{Hxc}}\left[\rho(\mathbf{r},t)\right]\right]|\psi_{n\mathbf{k}}(\mathbf{r},t)\rangle. \tag{4}$$

The second and third terms on the right-hand side are the atomic potential and the Hartree-exchange-correlation potential, respectively. Further details of this method are given in the Method section. The time profile of the cell-averaged current density can be calculated from the mechanical momentum expectation of time-evolving KS wavefunctions $|\psi_{n\mathbf{k}}(\mathbf{r},t)\rangle$:

$$\mathbf{J}(t) = -\frac{e}{Am} \sum_{n\mathbf{k}} f_{n\mathbf{k}} \langle \psi_{n\mathbf{k}}(\mathbf{r},t) | \hat{\pi} | \psi_{n\mathbf{k}}(\mathbf{r},t) \rangle, \tag{5}$$

where $A$ is the area of the 2D unit cell. The mechanical momentum operator is defined previously as $\hat{\pi} = \hat{\mathbf{p}} + \frac{e}{c}\mathbf{A}(t) + \frac{im}{\hbar}\left[\hat{V}_{NL}, \hat{\mathbf{r}}\right]$ with the nonlocal pseudopotential [8].

We trace the real-time dynamics of the transverse current density dynamics of the strained monolayer h-BN (B-N bond lengths are elongated by 0.2 Å in the *y*-direction), which has C$_{2v}$ point group symmetry (see Fig. 1b). The system is driven by the *x*-polarized electric field $\mathbf{E}(t) = f(t)E_0 \cos(\omega_0 t)\hat{\mathbf{x}}$, which is perpendicular to the mirror plane $\mathbf{M}_x$. We set $E_0 = 0.005$ V/Å and $\hbar\omega_0 = 2.89$ eV, which is 0.8 of the bandgap energy. Once the real-time profile of the transverse current $\mathbf{J}_\perp(t)$ is obtained within the [20 fs, 60 fs] range (the brown lines in Fig. 1c), we take its Fourier transformation $|F[\mathbf{J}_\perp(t)]|^2$ up to the fourth harmonic order of the applied frequency, as presented in Fig. 1d. The transverse current $J_\perp = J_y$ shows second-harmonic $2\omega_0$ oscillation (the brown peak in Fig. 1d). However, when the polarization of the applied field is rotated to the *y*-axis ($\mathbf{E} \parallel \mathbf{M}_x$) the transverse current $J_\perp = J_x$ vanishes (see the green lines in Figs. 1c-d). Originating from the odd symmetry of the Berry curvature with respect to the mirror plane, it has been previously shown that the Hall response vanishes when the applied light is parallel to the mirror plane [24,25]. Our rt-TDDFT results are consistent with this phenomenon and shows the polarization-dependency of the SHHE.

Next, we record the SHG yield of $|F[J_y(t)]|^2$ given in Fig. 1d and extend this result to a set of various light frequencies to construct a response spectrum in the sub-bandgap regime. The corresponding response here is the SHG current density response, which is $\omega\chi^{(2)}$ in terms of the SHG susceptibility $\chi^{(2)}$. For a 2D insulator, the imaginary part of the SHG susceptibility tensor $\chi^{(2)}$ is given by [13]

$$\chi^{(2)}(\omega) = \frac{2\pi i e^3}{\hbar^2} \int_{\mathbf{k}} \sum_{nmp} \frac{(f_{n\mathbf{k}} - f_{m\mathbf{k}}) v_{nm}^y [v_{mp}^x v_{pn}^x + v_{mp}^x v_{pn}^x]}{\omega_{mn}^3 (\omega_{mp} + \omega_{np})} [\delta(\omega_{mn} - 2\omega) - \delta(\omega_{mn} + 2\omega)]$$
$$= -\frac{8\pi e^3}{\hbar^2} \int_{\mathbf{k}} \mathrm{Im} \sum_{nmp} \frac{f_{n\mathbf{k}}(1 - f_{m\mathbf{k}}) v_{nm}^y v_{mp}^x v_{pn}^x}{\omega_{mn}^3 (\omega_{mp} + \omega_{np})} \delta(2\omega - \omega_{mn}) \quad \text{for } 2\hbar\omega > \varepsilon_{gap} > \hbar\omega,$$
(6)

where $\varepsilon_{gap}$ is the bandgap energy of the insulator and $\int_{\mathbf{k}} = \int \frac{d^2\mathbf{k}}{4\pi^2}$ stands for 2D Brillouin zone integration. The sum of each (*m*,*n*) and (*n*,*m*) pairs in the first line of Eq. (6) results in twice of the imaginary part of the expression inside the sum in the second line. Figs. 1e-f show the correspondence between the total SHG susceptibility spectra $|\omega\chi^{(2)}|$, and the rt-TDDFT current density results for two distinct point group symmetries of h-BN: $D_{3h}$ (pristine, see Fig. 2a) and $C_{2v}$ (strained, see Figs. 1b and 2b). Since the Fourier yield is a positive magnitude, we take the absolute value of the theoretical spectra. The rt-TDDFT calculations simply evolve the KS states by time-dependent Hamiltonian without considering any perturbative expansion. Therefore, this method is completely independent of the linear and nonlinear perturbation theories. Nevertheless, it produces quite consistent sub-bandgap SHG susceptibility spectra. The rt-TDDFT calculations confirm that the transverse response becomes significant when the light frequency exceeds one half the bandgap. This can be attributed to the SHG nature of the effect and the two-photon excitations in the nonlinear regime. We previously reported similar behavior in magnetic responses [26]. The present work can be considered as an electric response analogy to our previous work.

### C. Interband BCD and second harmonic Hall effect

The intraband BCD has been widely studied for NLHE of metallic Fermi surfaces [11,17] and the interband BCD has recently been defined for CPGE of insulating bands that are separated by the energy of the incident light [15,16]. Here, we propose a connection between the second harmonic NLHE of insulators and interband BCD. To reveal it we compare the SHG susceptibility tensor and the interband BCD in a few insulators that preserve TRS but lack the inversion symmetry. Here, we introduce the interband BCD of the band-edge states obtained from ground state KS wavefunctions for SHHE as follows

$$\mathbf{D}_{vc}(\omega) = \int_{\mathbf{k}} \frac{\partial \Omega_{vc}}{\partial \mathbf{k}} \Theta(2\omega - \omega_{cv}), \tag{7}$$

where $v$ = VBM, $c$ = CBM, and $\Theta$ is the Heaviside step function [16]. This is the interband BCD between valence band maxima (VBM) and conduction band minima (CBM) only. But one can easily define $\mathbf{D}_{nm}(\omega)$ for any valence band $n$ and conduction band $m$.

Throughout the NLHE, CPGE and SHHE, all the effect of the material's symmetry can be discussed in terms of the corresponding BCD. For example, when the system has reflection symmetry with respect to the $M_x$ mirror plane, the derivatives of both the intraband and interband Berry curvature must keep the relations $\left.\frac{\partial \Omega}{\partial k_y}\right|_{(-k_x,k_y)} = -\left.\frac{\partial \Omega}{\partial k_y}\right|_{(k_x,k_y)}$ and $\left.\frac{\partial \Omega}{\partial k_x}\right|_{(-k_x,k_y)} = \left.\frac{\partial \Omega}{\partial k_x}\right|_{(k_x,k_y)}$ because the Berry curvature is an odd function under $M_x$ mirror reflection, $\Omega(-k_x,k_y) = -\Omega(k_x,k_y)$. This indicates that the interband BCD, given in Eq. (7), is a pseudo-vector and must align along the $x$-axis, perpendicular to the mirror plane $M_x$. The presence of additional mirror planes vanishes the BCD [11,12,15,27]. The step function dictates that the $x$-component of $\mathbf{D}_{vc}(\omega)$ also vanishes when the light frequency is below $0.5\varepsilon_{gap}$.

Now, we decompose the SHG susceptibility tensor given in Eq. (6) into two parts: $\chi^{(2)} = \chi^{(2)}_{\text{2-band}} + \chi^{(2)}_{\text{3-band}}$, where $\chi^{(2)}_{\text{2-band}}$ counts the terms only when $p = m$ or $p = n$, and $\chi^{(2)}_{\text{3-band}}$ stands for all the other terms. The two terms can be explicitly written as

$$\chi^{(2)}_{\text{2-band}}(\omega) = \frac{8\pi e^3}{\hbar^2} \int_{\mathbf{k}} \text{Im} \sum_{nm} \frac{f_{n\mathbf{k}}(1-f_{m\mathbf{k}})v^y_{nm}v^x_{mn}}{\omega^4_{mn}} \Delta v^x_{mn} \delta(2\omega - \omega_{mn}), \qquad (8)$$

$$\chi^{(2)}_{\text{3-band}}(\omega) = -\frac{8\pi e^3}{\hbar^2} \int_{\mathbf{k}} \text{Im} \sum_{\substack{nmp \\ p \neq n \neq m}} \frac{f_{n\mathbf{k}}(1-f_{m\mathbf{k}})v^y_{nm}v^x_{mp}v^x_{pn}}{\omega^3_{mn}(\omega_{mp} + \omega_{np})} \delta(2\omega - \omega_{mn}), \qquad (9)$$

where $\Delta v^x_{mn} = v^x_{mm} - v^x_{nn}$ are the velocity expectation differences between the bands $m$ and $n$ at the Bloch point $\mathbf{k}$. Using the definition of the interband Berry curvature and the identity $\Delta v^x_{mn} = \frac{d\omega_{mn}}{dk_x}$, one can reduce $\chi^{(2)}_{\text{2-band}}$ to the summation of interband BCD for all occupied ($n$) and unoccupied ($m$) bands as

$$\chi^{(2)}_{\text{2-band}}(\omega) = \frac{\pi e^3}{\hbar^2 \omega^2} \sum_{nm} (\mathbf{D}_{nm} \cdot \hat{\mathbf{x}}) \approx \frac{\pi e^3}{\hbar^2 \omega^2} (\mathbf{D}_{vc} \cdot \hat{\mathbf{x}}), \qquad (10)$$

where we approximate that the main contributing term is the interband BCD of band-edge states, $\mathbf{D}_{vc}$, in the sub-bandgap regime. This expression shows the relation between the SHG signals and the interband BCD. The detailed derivation of Eq. (10) is given in supplementary material (SM). Since $\mathbf{D}_{vc}$ is a pseudo-vector, $\chi^{(2)}_{\text{2-band}}$ also preserves the same symmetry constraints and vanishes in the multiple reflection symmetries. On the other hand, $\chi^{(2)}_{\text{3-band}}$ cannot be reduced to such a pseudo-vector form and it is also even under the mirror reflection by even number of $v^x$ integrations. Thus, $\chi^{(2)}_{\text{3-band}}$ is not required to vanish by additional mirror planes in three-fold symmetric point groups.

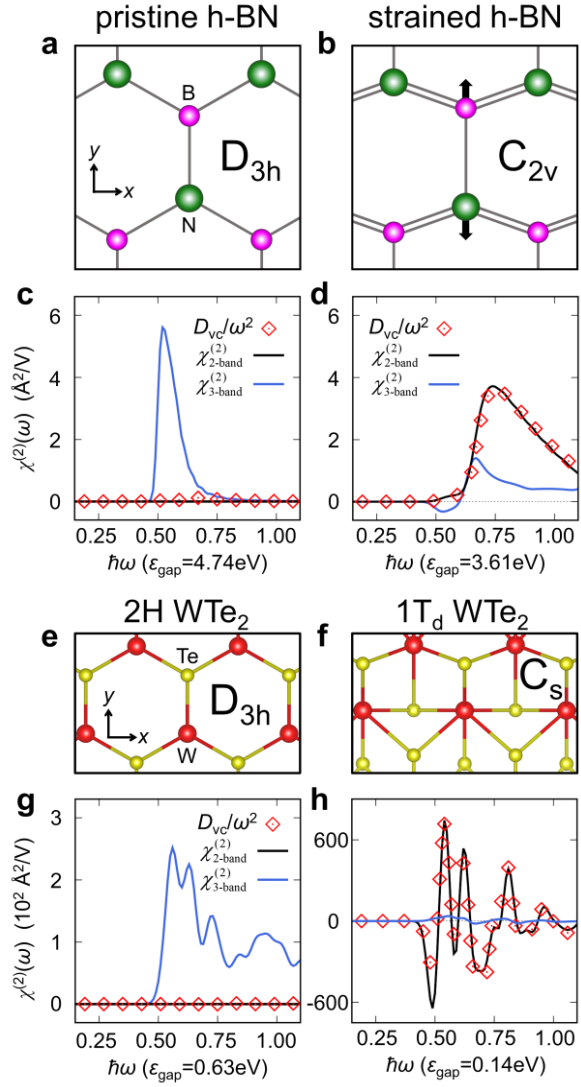

FIG. 2. **The correspondence between SHG susceptibility and the interband BCD. a** and **b** are the top views of three-fold symmetric $D_{3h}$ and two-fold symmetric $C_{2v}$ h-BN monolayer systems, respectively. The single (double) bond lines in **c** represent the long (short) bonds of the strained $C_{2v}$ system. **c** (**d**), the comparison of 2-band and 3-band transition terms of SHG susceptibility spectra with the interband BCD of the band-edge states of h-BN with $D_{3h}$ ($C_{2v}$) point group. **e-f** are same for monolayer WTe$_2$. **e**, **g** are for 2H phase of WTe$_2$ with $D_{3h}$ point group and **f**, **h** are for 1T$_d$ phase of WTe$_2$ with two-fold symmetric $C_s$ point group.

The comparison of $\chi^{(2)}_{\text{2-band}}$, $\chi^{(2)}_{\text{3-band}}$, and $\mathbf{D}_{vc}/\omega^2$ for three-fold and two-fold symmetric point groups is given in Fig. 2. The quantities are obtained from the ground state KS wavefunctions of the pristine ($D_{3h}$) and the strained ($C_{2v}$) h-BN monolayers (see Figs. 2a-d) and the 2H phase (D$_{3h}$) and the 1T$_d$ phase (C$_s$) of WTe$_2$ monolayers (see Figs. 2e-h). The correspondence between $\chi^{(2)}_{\text{2-band}}$ and $\mathbf{D}_{vc}/\omega^2$ is clearly seen in Figs. 3c, 3d, 3g, and 3h, which supports our approximation in Eq. (10). The dipoles $\mathbf{D}_{nm}$ between other bands are negligible, and thus, the band edge states are very representative part of the total BCD. In the three-fold D$_{3h}$ point group there are three vertical mirror planes, hence the BCD, and correspondingly $\chi^{(2)}_{\text{2-band}}$ vanishes as seen in Figs. 2c and 2g. However, such a symmetry restriction is not valid for $\chi^{(2)}_{\text{3-band}}$. Therefore, the SHG response solely comes from $\chi^{(2)}_{\text{3-band}}$ when the noncentrosymmetric insulator has more than one mirror planes. This response has recently been explained in terms of skew scattering through Berry curvature triple, a higher-order moment of the Berry curvature [28]. On the other hand, BCD (or $\chi^{(2)}_{\text{2-band}}$) becomes non-zero in the C$_{2v}$ and C$_s$ point groups, even further it is larger than $\chi^{(2)}_{\text{3-band}}$ and dominates overall SHG response (see Figs. 2d and 2h). Comparing the denominators of Eqs. (8) and (9), it is obvious that $\omega_{mn} < (\omega_{mp} + \omega_{np})/2$ and $\chi^{(2)}_{\text{3-band}} < \chi^{(2)}_{\text{2-band}}$ when $\omega_{mn}$ is $\omega_{cv}$. Hence, the dominance of the BCD is expected to be more significant when the bandgap is small and when the band-edge states are well separated from other bands. Comparison of Figs. 2d and 2h clearly shows that the WTe$_2$ has four order of magnitude larger response than h-BN, and it is almost entirely coming from BCD contribution. This means that the SHG signals of the narrow-gap insulators with the single mirror plane mainly originates from the interband BCD of the band-edge states. Thus, the corresponding SHHE current density response can be formulated in terms of BCD as

$$\mathbf{J}_\perp(t) = -\frac{2\pi e^3}{\hbar^2 \omega} \hat{\mathbf{z}} \times \mathbf{E}(\mathbf{D}_{vc} \cdot \mathbf{E}), \tag{11}$$

which is general form of Eq. (10) for any light polarization and explains the polarization dependency of the SHHE response observed in Figs. 1c-d. This correlation suggests a realizable method to detect the material's Berry curvature properties by using SHG spectroscopy. By aid of the simplified model Hamiltonian calculations, we present a comprehensive discussion on the correspondence between material symmetry and Berry curvature distribution, BCD spectra and SHG responses in SM. Those results further suggest ways for adjustment of the SHG signals by controlling the material symmetry and extracting the Berry curvature distribution from a set of SHG spectra.

**D. The SHHE under strong fields: inclusion of Berry curvature multipoles**

We now analyze the behavior of the induced Hall current under strong fields $\mathbf{E}(t) = E_0 \operatorname{Re}[e^{i\omega t}]\hat{\mathbf{x}}$ with large $E_0$. Assuming the contribution of only 2-band term dominates, the second-harmonic generation current in the $y$-direction can be obtained from Eq. (11) as

$$J_y(t) = -\frac{2\pi e^3}{\hbar^2 \omega} E_0^2 \operatorname{Re}[e^{i2\omega t}] \int_\mathbf{k} \frac{\partial \Omega_{vc}}{\partial k_x} \Theta(2\omega - \omega_{cv}). \tag{12}$$

Application of non-perturbative strong fields induce intraband motion of electrons together with interband transitions. The intraband motion of the carriers in the momentum space can be described by acceleration theorem as $\mathbf{k} \to \mathbf{k} + \frac{e}{\hbar c}\mathbf{A}(t)$ with the corresponding vector potential $\mathbf{A}(t) = -\frac{c}{\omega} E_0 \operatorname{Im}[e^{i\omega t}]\hat{\mathbf{x}}$. The shift in the momentum is reflected in the BCD term as $\frac{\partial \Omega_{vc}}{\partial k_x} \to \frac{\partial \Omega_{vc}}{\partial k_x}\bigg|_{\mathbf{k}+\Delta k\hat{\mathbf{x}}}$ with $\Delta k(t) = -\frac{e}{\hbar \omega} E_0 \operatorname{Im}[e^{i\omega t}]$. Here, we assume that the step function is related to the initial position of excited carriers in the BZ, and thus, we do not apply same shift to it. Now, we expand the BCD with respect to $\Delta k$ and sum for each Kramers pairs (**k**, −**k**) as:

$$\frac{\partial \Omega_{vc}}{\partial k_x}\bigg|_{\mathbf{k}+\Delta k\hat{\mathbf{x}}} + \frac{\partial \Omega_{vc}}{\partial k_x}\bigg|_{-\mathbf{k}+\Delta k\hat{\mathbf{x}}} = 2\frac{\partial \Omega_{vc}}{\partial k_x} + 2\frac{1}{2!}\frac{\partial^3 \Omega_{vc}}{\partial k_x^3}(\Delta k)^2 + 2\frac{1}{4!}\frac{\partial^5 \Omega_{vc}}{\partial k_x^5}(\Delta k)^4 + \ldots, \tag{13}$$

where all the odd order terms vanish by the oddness of $\Omega_{vc}$ in the presence of TRS. By collecting the zero-frequency terms of $(\Delta k)^2$, $(\Delta k)^4$, etc., we can rewrite Eq. (12) as

$$J_y(t) = -\text{Re}[e^{i2\omega t}]\left\{\frac{2\pi e^3}{\hbar^2 \omega}E_0^2 \mathbf{D}_{vc}^{(2)} + \frac{\pi e^5}{\hbar^4 \omega^3}E_0^4 \mathbf{D}_{vc}^{(4)} + \frac{7\pi e^7}{192\hbar^6 \omega^5}E_0^6 \mathbf{D}_{vc}^{(6)} + \ldots\right\}\cdot \hat{\mathbf{x}} \quad (14)$$

where $\mathbf{D}_{vc}^{(2)} = \mathbf{D}_{vc}$ and $\mathbf{D}_{vc}^{(n)}(\omega) = \int_{\mathbf{k}}\frac{\partial^{(n-1)}\Omega_{vc}}{\partial \mathbf{k}^{(n-1)}}\Theta(2\omega - \omega_{cv})$ are interband BCD and interband Berry curvature $n$-pole of band-edge states, respectively. A similar definition of the Berry curvature multipoles has recently been discussed for the higher-order nonlinear responses of metals [29]. The SHG response consists of high-order contributions through Berry curvature multipoles in the strong field regime, which can be considered as the non-perturbative regime. Thus, it is not quadratically increasing by the field strength. Note that the contributions given in Eq. (14) do not represent the entire physics of the nonlinear current. Here, we consider application of the acceleration theorem and corresponding expansion of the current expectation obtained from the 2-band term of the second-order perturbation theory, $\chi_{\text{2-band}}^{(2)}$. On the other hand, the expansion of the semi-classical anomalous current, which corresponds to the first-order susceptibility $\chi^{(1)}$, also produces similar terms associated with Berry curvature multipoles (see SM). Together with these, the intraband motion of electrons under strong fields requires the expansion of higher-order perturbation currents, $\chi^{(n)}$ as well. They consist of many diverse terms, and thus, only some of their expansion terms (2-band terms) could lead to Berry curvature multipoles, while others are not, such as the 3-band term of the second order current, $\chi_{\text{3-band}}^{(2)}$. To describe the correct physics, all contributions should be considered. But, here, we only claim that the direct correspondence between the SHG responses and BCD is limited for weak fields, and strong fields induce higher-order contributions with some portion originated from the Berry curvature multipoles.

We search for a strong field limit and analyze how real-time SHG responses deviates from the second-order perturbation theory. Figs. 3a-b show the same calculation results with Figs. 1c-d for a 100 times larger field strength. While the weak field produces only SHG (see Fig. 1d), the strong field clearly shows up to 10$^{\text{th}}$ harmonic generation (see Fig. 3b). To find the field strength that can generate *n*-th harmonic response and to compare the trend of HHG yields we plot Fig. 3c. For the fixed light frequency of $\hbar\omega_0 = 0.6\varepsilon_{\text{gap}}$ we vary the field strength in the window of [0.001 V/Å, 1.000 V/Å] and record the HHG yields. The SHG response is not visible for the weak fields such as 0.001 V/Å and less. Similarly, 4HG, 6HG and 8HG yield peaks becomes significant when the field strength exceeds 0.02 V/Å, 0.05 V/Å and 0.1 V/Å, respectively. The SHG yield shows clear quadratic increase ($\propto E_0^2$) with the strengths from 0.002 V/Å to 0.4 V/Å. However, this behavior drops for stronger fields, which indicates that the system reaches the non-perturbative limit. Specifically at 0.5 V/Å, all the peak yields are comparable as seen in Fig. 3c. Thus, it can be a good example to show effect of high-order contributions in SHG.

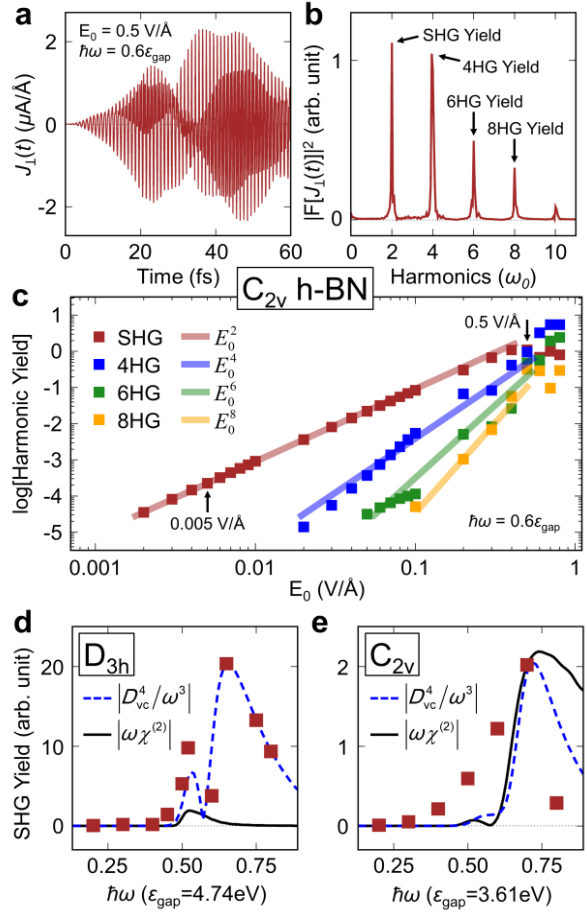

**FIG. 3. The SHHE response to strong fields and contribution of the Berry curvature multipoles. a**, The real-time profile of the transverse current density response of $C_{2v}$ h-BN obtained by rt-TDDFT calculations. The field strength is 100 times larger than that of Fig. 1c. **b**, The Fourier transformation spectra of the current density given in **a** for the time interval of [20 fs,60 fs]. The x-axis is scaled with the applied field frequency, $\hbar\omega_0 = 0.6\varepsilon_{gap}$. The SHG and HHG yields are depicted with arrows. **c**, The SHG and HHG yields for various field strengths. Both x and y-axes are log scaled. Arrows show the field strengths used in Figs. 1c-f (0.005 V/Å) and Figs. 3a-b, 3d-e (0.5 V/Å). The transparent red, blue, green, and yellow lines represent the perturbative power rules for SHG, 4HG, 6HG and 8HG yield, respectively. **d** and **e** are the comparison of the theoretical SHG susceptibility spectra also shown in Figs. 1e-f and the SHG yields for $D_{3h}$ and $C_{2v}$ h-BN, respectively. The light frequency is scaled with the bandgap of each structure. The brown filled squares are the SHG yields of the transverse current as given in **c** for 0.5 V/Å. The black lines are the same as Figs. 1e-f with a larger scaling factor that fit the SHG susceptibility to the maximum of SHG yield in **e**. The green dashed lines are Berry curvature 4-pole divided by $\omega^3$ that represents the contribution of high-order corrections to SHG response. The lines are fitted to the maximum of the SHG yield data in both **d** and **e**.

As done in Figs. 1e-f, we compare the theoretical perturbative SHG susceptibility and the SHG yields obtained from the rt-TDDFT calculations for the strong field, 0.5 V/Å. We first adjust the theoretical curve (black lines) to the maximum of SHG yield (brown squares) of $C_{2v}$ h-BN in Fig. 3e. And then by using the same constant we draw the theoretical curve in Fig. 3d. The same fitting method was used in Figs. 1e-f and both geometries show a great consistency between theory and calculation. But now, the field is over the perturbative regime, and the SHG yields deviate from the theoretical expectations. Compared with the weak field (see Figs. 1e-f), especially $D_{3h}$ geometry show an enhanced response to the strong field (see Fig. 3d), which can be attributed to Eq. (14). The three-fold symmetry vanishes the interband BCD ($\mathbf{D}_{vc}^{(2)}$); however, multipoles ($\mathbf{D}_{vc}^{(4)}$, $\mathbf{D}_{vc}^{(6)}$, …) are allowed and their effect becomes significant under the strong field. As seen in Fig. 3d, the overall shape of the SHG yield spectrum mostly consistent with the Berry curvature 4-pole, with a small contribution from the 3-band transition term as discussed above. On the other hand, $C_{2v}$ structure has differences in the shape of SHG yield spectra for weak and strong fields. Moreover, addition of $\mathbf{D}_{vc}^{(4)}$ also does not ameliorate the deviation. Emergence of the differences and responses around half the gap can be originated from the high-order terms, which do not correspond to the Berry curvature multipoles.

## IV. CONCLUSION

We investigated the second harmonic Hall responses of inversion-broken TR-symmetric insulators in the framework of the real-time propagation of time-dependent density functional theory. The optical response spectra of SHG signals can be accurately obtained from rt-TDDFT calculations in the sub-bandgap frequency regime. We closely examine the SHG susceptibility tensor and found that the response is dominated by the interband BCD, when the insulator preserves single mirror symmetry. The correspondence between second harmonic Hall effect and interband BCD suggests the use of SHG spectroscopy analysis to detect insulators' Berry curvature distribution. The SHG response to strong fields contains more information about the band structure geometry through Berry curvature multipoles.

## V. METHOD

**Real-time time-dependent density functional theory calculations**

The ground-state electronic structure was obtained by a standard DFT calculation using the Quantum Espresso package [30]. The preferred exchange and correlation functional was the Perdew–Burke–Ernzerhof (PBE)-type generalized gradient approximation functional [31]. The scalar-relativistic norm-conserving pseudopotentials were exploited to describe the atomic potentials [32]. The cutoff energy was set as 1088 eV. The primitive unit cells with a vacuum slab of 15 Å were used for the monolayers. The Brillouin zone integration was carried out following a Monkhorst–Pack scheme of 24×24×1 mesh for time-dependent calculations, and 60×60×1 mesh for static BCD calculations, excluding any symmetry operation. For the computations of the time propagations, we used the plane-wave package developed by our group [33], employing the Suzuki–Trotter time propagation scheme [34]. The time interval for real-time evolution was set at 0.048 fs.

## ACKNOWLEDGEMENT

This work was supported by an NRF grant (No. NRF-2019R1A2C2089332). We thank Prof. Jungwoo Kim from Incheon National University and Kyoung-Whan Kim from Korea Institute of Science and Technology for productive discussions

## Data and code availability

The calculated numerical data and the Fortran code used to perform the calculations that support our study are available from the corresponding author upon reasonable request.

## Author contributions

M.S.O. performed ab initio calculations, developed the model Hamiltonian code, and analyzed the data; M.S.O. and N.P. edited the first draft of the manuscript. All authors discussed and analyzed the results and contributed to and commented on the manuscript.

## Competing interests

The authors declare that they have no competing interests.

# SUPPLEMENTARY MATERIAL

## A. Derivation of Berry curvature dipole formalism from SHG susceptibility

**Proof 1**. We start from the two-band transition part of the SHG susceptibility tensor

$$\chi^{(2)}_{2\text{-band}}(\omega) = \frac{8\pi e^3}{\hbar^2} \int \frac{d^2\mathbf{k}}{4\pi^2} \operatorname{Im} \sum_{nm} \frac{f_{n\mathbf{k}}(1-f_{m\mathbf{k}})v^y_{nm}v^x_{mn}}{\omega^4_{mn}} \Delta v^x_{mn} \delta(2\omega - \omega_{mn}), \tag{1}$$

where $\Delta v^x_{mn} = v^x_{mm} - v^x_{nn}$ as given in the main text and $f_{n\mathbf{k}}(1-f_{m\mathbf{k}})$ requires $n$ to be a valence band and $m$ to be a conduction band. By introducing the interband Berry curvature $\Omega_{nm}(\mathbf{k}) = 2\operatorname{Im}\frac{f_{n\mathbf{k}}(1-f_{m\mathbf{k}})v^y_{nm}v^x_{mn}}{\omega^2_{mn}}$, and using the identity $\Delta v^x_{mn} = \frac{d(\omega_{mn})}{dk_x}$, we have

$$\begin{aligned}\chi^{(2)}_{2\text{-band}}(\omega) &= \frac{4\pi e^3}{\hbar^2} \sum_{nm} \int_{\mathbf{k}} \frac{\Omega_{nm}}{\omega^2_{mn}} \frac{d(\omega_{mn})}{dk_x} \delta(2\omega-\omega_{mn}) \\ &= -\frac{\pi e^3}{\hbar^2 \omega^2} \sum_{nm} \int_{\mathbf{k}} \Omega_{nm} \frac{d\left(\Theta(2\omega-\omega_{mn})\right)}{dk_x}\end{aligned} \tag{2}$$

where the delta function and step function are assumed as differentiable by holding the relation $\frac{d}{dk_x}\Theta(2\omega-\omega_{mn}) = -\delta(2\omega-\omega_{mn})\frac{d(\omega_{mn})}{dk_x}$, and the delta function in the first line dictates the frequency selection as $\omega_{mn} = 2\omega$. Now, we apply integration by parts

$$\int_\mathbf{k} \Omega_{nm} \frac{d\left(\Theta(2\omega-\omega_{mn})\right)}{dk_x} + \int_\mathbf{k} \frac{d(\Omega_{nm})}{dk_x} \Theta(2\omega-\omega_{mn}) = \Omega_{nm}\Theta(2\omega-\omega_{mn})\big|_{\text{BZ boundary}}, \tag{3}$$

where right hand side vanishes since both $\Omega_{nm}$ and $\omega_{mn}$ hold the periodic boundary conditions. Therefore, Eq. (2) becomes

$$\chi^{(2)}_{2\text{-band}}(\omega) = \frac{\pi e^3}{\hbar^2 \omega^2} \sum_{nm} \int_\mathbf{k} \frac{d\Omega_{nm}}{dk_x} \Theta(2\omega-\omega_{mn}). \tag{4}$$

The interband BCD for SHHE is defined as $\mathbf{D}_{nm}(\omega) = \int_\mathbf{k} \frac{\partial \Omega_{nm}}{\partial \mathbf{k}} \Theta(2\omega - \omega_{mn})$, and thus we reach

Eq. (10) of the main text

$$\chi^{(2)}_{\text{2-band}}(\omega) = \frac{\pi e^3}{\hbar^2 \omega^2} \sum_{nm} (\mathbf{D}_{nm} \cdot \hat{\mathbf{x}}). \tag{5}$$

**Proof 2.** Again, we start from the two-band term of the SHG susceptibility tensor

$$\chi^{(2)}_{\text{2-band}}(\omega) = \frac{8\pi e^3}{\hbar^2} \int \frac{d^2\mathbf{k}}{4\pi^2} \,\text{Im} \sum_{nm} \frac{f_{n\mathbf{k}}(1-f_{m\mathbf{k}}) v^y_{nm} v^x_{mn}}{\omega^4_{mn}} \Delta v^x_{mn} \delta(2\omega - \omega_{mn}), \tag{6}$$

and now we split the integration into two parts depending on the sign of $\Delta v^x_{mn}$ in the Brillouin zone as

$$\chi^{(2)}_{\text{2-band}}(\omega) = \frac{e^3}{\pi\hbar^2} \sum_{nm} \int dk_y \left\{ \begin{array}{l} \displaystyle\int\limits_{\omega_{mn}>0}^{(\Delta v^x_{mn}>0)} d(\omega_{mn}) \frac{\Omega_{nm}}{\omega^2_{mn}} \delta(2\omega-\omega_{mn}) \\ \displaystyle -\int\limits_{\omega_{mn}>0}^{(\Delta v^x_{mn}<0)} d(\omega_{mn}) \frac{\Omega_{nm}}{\omega^2_{mn}} \delta(2\omega-\omega_{mn}) \end{array} \right\}, \tag{7}$$

Since $\omega_{mn}$ is an even function of $k_x$, $\Delta v^x_{mn}$ is odd and its sign should be considered in the integrations.

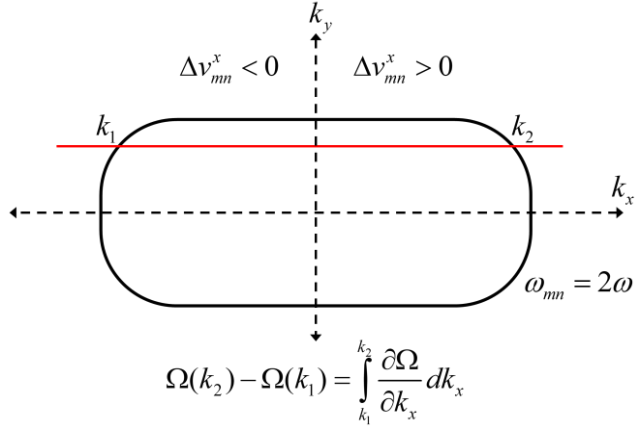

**FIG. S1. A sample illustration of the integration contour.** The solid black line is the closed contour $\omega_{mn} = 2\omega$. If we assume that the band gap is at the origin and the energy gap $\omega_{mn}$ is monotonically increasing as k moves away from the origin, then the sign of $\Delta v^x_{mn}$ is same with the sign of $k_x$. A line parallel to the x-axis (the red line) intersects the loop at the points whose x-components are $k_1$ and $k_2$, which satisfy $k_1 = -k_2$.

The integrations with delta functions in Eq. (7) result in line integrals of $k_y$:

$$\chi^{(2)}_{\text{2-band}}(\omega) = \frac{e^3}{\pi\hbar^2(2\omega)^2} \sum_{nm} \left\{ \int_{\omega_{mn}=2\omega}^{(\Delta v^x_{mn}>0)} dk_y \Omega_{nm} - \int_{\omega_{mn}=2\omega}^{(\Delta v^x_{mn}<0)} dk_y \Omega_{nm} \right\}, \quad (8)$$

where closed contours $\omega_{mn} = 2\omega$ are divided into two parts. For each $k_y$ value within the contour, we assume a line parallel to the $x$-axis (see Fig. S1). The two points ($k_1$ and $k_2$) that are the intersections of the contour and the line passing through $k_y$ must belong to different integrals given in Eq. (8). Thus, the quantity we calculate in Eq. (8) would be the difference in Berry curvature of the two ends of the line, $\Omega_{nm}(k_2) - \Omega_{nm}(k_1)$. This quantity is identical to the first order derivative of the Berry curvature with respect to $k_x$ integrated over the line segment that connects the contour (the red line that connects $k_1$ and $k_2$), $\int_{k_1}^{k_2} \frac{\partial \Omega_{nm}}{\partial k_x} dk_x$. When this process is repeated for every $k_y$, we reach the surface integration within the contour as follows:

$$\chi^{(2)}_{\text{2-band}}(\omega) = \frac{\pi e^3}{\hbar^2 \omega^2} \sum_{nm} \int \frac{d^2\mathbf{k}}{4\pi^2} \frac{\partial \Omega_{nm}}{\partial k_x} \Theta(2\omega - \omega_{mn}), \quad (9)$$

which is identical to Eq. (5).

### B. The two-band model Hamiltonian and corresponding time evolution

We study the Haldane model, which is a suitable model for h-BN and convenient for adjusting the TRS and the topology [35]. The 2×2 Hamiltonian matrix is

$$\hat{H}(\mathbf{k}) = \left(h_x(\mathbf{k}), h_y(\mathbf{k}), m + \mu(\mathbf{k})\right) \cdot \hat{\boldsymbol{\sigma}}, \quad (10)$$

where $\hat{\boldsymbol{\sigma}}$ are the Pauli matrices in the sublattice space of a hexagonal unit cell. The momentum-dependent structural functions are defined as $h_x(\mathbf{k}) - ih_y(\mathbf{k}) = -t_H \sum_{i=1}^{3} e^{i\mathbf{k}\cdot\boldsymbol{\tau}_i}$ and $\mu_z(\mathbf{k}) = \frac{4}{3a^2} \lambda \sum_{i=1}^{3} \sin(\mathbf{k} \cdot \upsilon_i) d_i$. Here, $t_H$ is the hopping integral, $a$ is the lattice length, $\tau_i$ are the position vectors pointing to the nearest-neighbor atoms, $\upsilon_i = \tau_j - \tau_k$, and $d_i = (\tau_j \times \tau_k)_z$. Breaking the TRS and the inversion symmetry results in parameters $\lambda$ and $m$, respectively. We set the parameters as $(\lambda, m) = (0.0 \text{ eV}, 1.5 \text{ eV})$ for a TR-symmetric insulator, and $(t_H, a) = (2.50 \text{ eV}, 2.46 \text{ Å})$. Thus, the bandgap $\varepsilon_{gap}$ is 3.0 eV.

The Haldane model originally preserves three-fold mirror symmetry including $M_x$ mirror plane ($D_{3h}$ point group). The symmetry is artificially lowered by an extension of the bond lengths along the $y$-axis ($C_{2v}$ point group). This extension is reflected in the model by $h_x(\mathbf{k}) - i h_y(\mathbf{k}) \to -\sum_{i=1}^{3} t_H^i e^{i\mathbf{k} \cdot \tilde{\tau}_i}$, where modified site-dependent hopping terms are $t_H^i = t_H e^{-3(|\tilde{\tau}_i| - a)/a}$ and extended bond vectors are $\tilde{\tau}_i = \tau_i - d\hat{\mathbf{y}} = \tau_i - (0.1 \text{ Å})\hat{\mathbf{y}}$ with $d$ being the extension of the bond length in the $y$-direction. The band structure and the Berry curvature of both symmetries are given in Fig. S2.

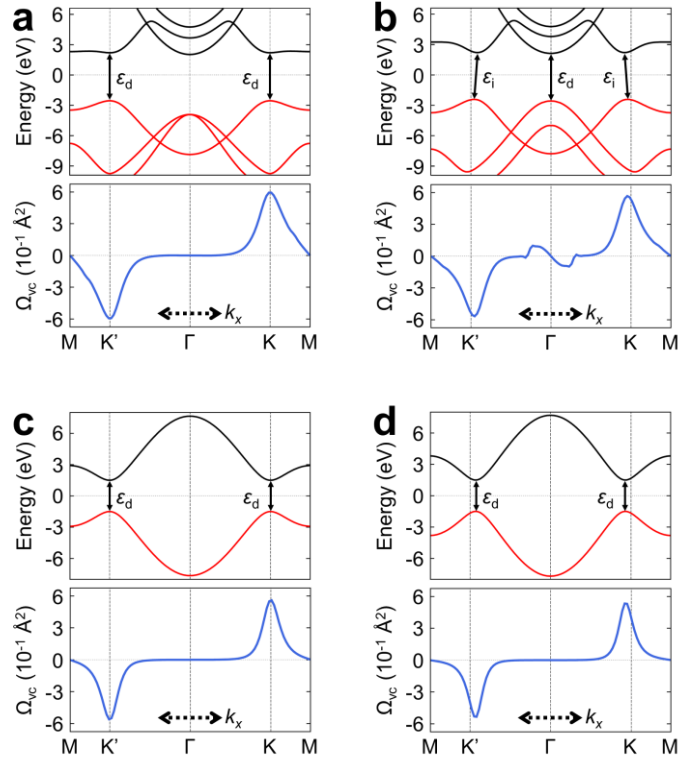

**FIG. S2. The geometry, the band structure and the Berry curvature of pristine and strained h-BN. a** and **b** show the band structure (upper panel) with the interband Berry curvature (lower panel) along the $k_x$ axis of pristine and strained h-BN monolayer obtained by DFT calculations, respectively. The direct bandgap in **a** is $\varepsilon_d$ = 4.74 eV, the direct and indirect bandgaps in **b** are $\varepsilon_d$ = 4.70 eV and $\varepsilon_i$ = 4.62 eV, respectively. **c** and **d** are same results obtained by two-band model calculations of TR-symmetric trivial insulator. The direct bandgaps are same in **c** and **d**, $\varepsilon_d$ = 3.00 eV. The red (black) lines represent occupied (unoccupied) states.

The most significant impact of the symmetry lowering on the band structure and the Berry curvature is the deviation of maxima points from K and K' valleys. This deviation induces non-vanishing interband BD. The model calculations mostly exhibit this behavior. DFT calculations additionally shows the indirect bandgaps near valleys, the lifting of valence band maximum at Γ point, and the induction of additional Berry curvature around Γ point consequently. For the sake of convenience, the Berry curvature, the interband BCD and the SHG susceptibility tensor can be reformulated for the two-band model that consists of a valence band (*v*) and a conduction band (*c*) as

$$\Omega_{vc}(\mathbf{k}) = 2\,\mathrm{Im}\frac{v_{vc}^{y} v_{cv}^{x}}{\omega_{cv}^{2}}, \tag{11}$$

$$D_{vc}(\omega) = \int_{\mathbf{k}} \frac{\partial \Omega_{vc}}{\partial k_{x}} \Theta(2\omega - \omega_{cv}), \tag{12}$$

$$\chi^{(2)}(\omega) = \frac{8\pi e^{3}}{\hbar^{2}} \int_{\mathbf{k}} \mathrm{Im}\frac{v_{vc}^{y} v_{cv}^{x}}{\omega_{cv}^{4}} \Delta v_{cv}^{x} \delta(2\omega - \omega_{cv}), \tag{13}$$

and as there is not any possible 3-band term, they hold the relation

$$\chi^{(2)}(\omega) = \frac{\pi e^{3}}{\hbar^{2} \omega^{2}} D_{vc}(\omega). \tag{14}$$

To work on the time-dependent model Hamiltonian calculations, we apply the $x$-polarized electric field $\mathbf{E}(t) = f(t) E_{0} \cos(\omega_{0} t)\hat{\mathbf{x}}$, construct the time-dependent Hamiltonian $\hat{H}(\mathbf{k},t) = \hat{H}(\mathbf{k} + \frac{e}{\hbar c}\mathbf{A}(t))$ through the Peierls substitution, and evolve the two-component wavefunctions numerically using Schrödinger's equation $|\psi_{n\mathbf{k}}(t+\Delta t)\rangle = e^{-i\hat{H}(\mathbf{k},t)\Delta t/\hbar}|\psi_{n\mathbf{k}}(t)\rangle$, where the time interval for the evolution is $\Delta t = 0.005$ fs.

### C. Estimation of the Berry curvature distribution by the SHHE spectra

We discuss the correlation between the Berry curvature distribution and the second harmonic Hall current responses in the framework of the two-band model Hamiltonian. By varying the parameter $d$, which determines the symmetry lowering from D$_{3h}$ to C$_{2v}$ by amount of the strain (bond length extension), and the parameter $m$, which stands for inversion breaking, we plot the band structures, the corresponding interband Berry curvatures $\Omega_{vc}(\mathbf{k})$, and the SHG susceptibility tensors $\chi^{(2)}(\omega)$ obtained from Eq. (14) in Fig. S3. In real experiment, $d$ and $m$ can be controlled by application of in-plane uniaxial strain and out-plane static electric field, respectively.

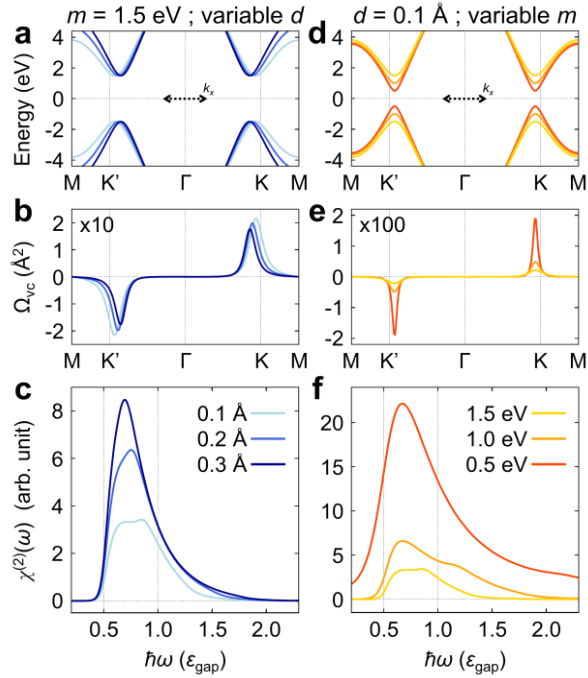

**FIG. S3. The effect of the Berry curvature distribution on SHHE. a**, **b**, and **c** are the band structures, the interband Berry curvatures, and the SHG susceptibility tensor spectra of the two-band model Hamiltonian, respectively, when the parameter $d$ varies among 0.1 Å (light blue lines), 0.2 Å (blue lines), and 0.3 Å (dark blue lines) with fixed $m$ = 1.5 eV. **d**, **e**, and **f** are same with **a**, **b**, and **c** for the parameter $m$ varies among 1.5 eV (yellow lines), 1.0 eV (orange lines), and 0.5 eV (dark orange lines) with fixed $d$ = 0.1 Å. The spectrum given in Fig. 1e corresponds to the parameters $m$ = 1.5 eV and $d$ = 0.1 Å.

When the strain in the *y*-direction becomes stronger (larger $d$) the band edge states deviate from the K and K' points in the $k_x$ direction more (Fig. S3a), and correspondingly the peak positions of the Berry curvature also shift in the same direction with a slight decrease in the peak amplitude (Fig. S3b). The effect of this shift is reflected in the response spectra as increasing the response amplitude with similar width (Fig. S3c). The amplitude increases since the BCD increases by symmetry lowering; the width keeps same since the bandgap does not change by broken mirror symmetry. On the other hand, for the materials with a smaller on-site energy difference (smaller $m$, less inversion-broken) the energy bandgap is smaller (Fig. S3d), but inversely the peak amplitude of the Berry curvature is much higher, and the peaks are
7

narrower compared with the materials with larger on-site energy difference (Fig. S3e). The narrower and higher Berry curvature peaks result in a wider response spectrum with larger amplitude, which can be mainly attributed to the difference in size of the bandgap (Fig. S3f). By analyzing the SHHE response spectra of two distinct systems, we can estimate a clear comparison between their Berry curvature distributions. Now we can suggest a method to detect the Berry curvature distribution of an insulator. A set of SHG spectra can be obtained under different conditions: application of uniaxial in-plane strain and out-plane electric field with various strengths. By comparing trend of the observed spectra with the theoretical spectral set, the best fitting corresponding Berry curvature distribution can be predicted. For this prediction, machine learning algorithms can be developed. We leave this point for the future studies. In addition, the control of material symmetry can be utilized to enhance the SHG signals in laboratory experiments.

### D. The semiclassical approach: effect of Berry curvature multipoles on SHHE

We consider the oscillating electric field of $\mathbf{E}(t) = E_0 \operatorname{Re}[e^{i\omega t}]\hat{\mathbf{x}}$, and assume that the semiclassical field allows substantial ratio of the two-photon process, leaving the hole and electron carriers in the band edge states satisfying $\varepsilon_m(\mathbf{k}) - \varepsilon_n(\mathbf{k}) = \hbar\omega_{mn}(\mathbf{k}) \leq 2\hbar\omega$. These states can be counted by the step function $\Theta(2\omega - \omega_{mn}(\mathbf{k}))$ in momentum space integrations. We can assume that the light first excites carriers, and then the readily excited carriers are subject to the semiclassical Bloch oscillation by same light field. The motion of the carriers in the momentum space can be described by Peierls substitution as $\mathbf{k} \to \mathbf{k} + \frac{e}{\hbar c}\mathbf{A}(t)$ with the corresponding vector potential $\mathbf{A}(t) = -\frac{c}{\omega}E_0 \operatorname{Im}[e^{i\omega t}]\hat{\mathbf{x}}$. And thus, the location of the carriers oscillates by the step function as $\Theta(2\omega - \omega_{mn}(\mathbf{k} + \Delta k \hat{\mathbf{x}}))$ with $\Delta k(t) = -\frac{e}{\hbar\omega} E_0 \text{Im}[e^{i\omega t}]$. The net current up to the first order can be found in terms of the relative excitonic velocity as $\mathbf{J} = \mathbf{J}^e + \mathbf{J}^h = -e\mathbf{v}^e + e\mathbf{v}^h = -e(\mathbf{v}^e - \mathbf{v}^h) = -e\mathbf{v}^{\text{rel}}$. The relative excitonic anomalous velocity between a hole carrying valence band $v$ and an electron carrying conduction band $c$ has been defined as $\mathbf{v}_\perp^{\text{rel}}(\mathbf{k}) = \dot{\mathbf{k}} \times [\mathbf{\Omega}_v(\mathbf{k}) - \mathbf{\Omega}_c(\mathbf{k})]$, where $\dot{\mathbf{k}} = \frac{e}{\hbar c}\dot{\mathbf{A}}(t) = -\frac{e}{\hbar}\mathbf{E}(t)$ [36]. If this concept is extended to a multi-band system, where a few bands contribute to the two-photon excitation, the total relative anomalous velocity can be obtained as $\mathbf{v}_\perp^{\text{rel}}(\mathbf{k}) = \dot{\mathbf{k}} \times \left[\sum_n f_{n\mathbf{k}} \mathbf{\Omega}_n(\mathbf{k}) - \sum_m (1-f_{m\mathbf{k}}) \mathbf{\Omega}_m(\mathbf{k})\right]$. Note that the total Berry curvature is identical to total interband Berry curvature

$$\sum_n f_{n\mathbf{k}} \Omega_n(\mathbf{k}) = 2\text{Im}\sum_{nm} f_{n\mathbf{k}}(1-f_{m\mathbf{k}})\frac{v_{nm}^y v_{mn}^x}{\omega_{mn}^2} + 2\text{Im}\sum_n \sum_{m\neq n} f_{n\mathbf{k}} f_{m\mathbf{k}} \frac{v_{nm}^y v_{mn}^x}{\omega_{mn}^2} \quad (15)$$
$$= \sum_{nm} \Omega_{nm}(\mathbf{k}),$$

$$\sum_m (1-f_{m\mathbf{k}})\Omega_m(\mathbf{k}) = 2\text{Im}\sum_{nm} f_{n\mathbf{k}}(1-f_{m\mathbf{k}})\frac{v_{mn}^y v_{nm}^x}{\omega_{mn}^2} + 2\text{Im}\sum_m \sum_{n\neq m}(1-f_{n\mathbf{k}})(1-f_{m\mathbf{k}})\frac{v_{nm}^y v_{mn}^x}{\omega_{mn}^2} \quad (16)$$
$$= -\sum_{nm} \Omega_{nm}(\mathbf{k}),$$

where the last terms in first lines of Eqs. (S10) and (S11) vanish since the sums include both $(m,n)$ and $(n,m)$ pairs, which results in $\text{Im } v_{mn}^y v_{nm}^x = \text{Im}(v_{nm}^y v_{mn}^x)^* = -\text{Im } v_{nm}^y v_{mn}^x$ for all $n$ and $m$ states [37-39]. By extracting the identity $2\sum_{nm} \Omega_{nm}(\mathbf{k}) = \sum_n f_{n\mathbf{k}}\Omega_n(\mathbf{k}) - \sum_m (1-f_{m\mathbf{k}})\Omega_m(\mathbf{k})$, the net transverse current of the electron and hole carriers at a time $t$ becomes

$$\mathbf{J}_\perp(t) = 2\frac{e^2}{\hbar}\mathbf{E}(t) \times \sum_{nm}\int \frac{d^2\mathbf{k}}{4\pi^2} \mathbf{\Omega}_{nm}(\mathbf{k})\Theta(2\omega - \omega_{mn}(\mathbf{k}+\Delta k\hat{\mathbf{x}})). \quad (17)$$

The Bloch oscillations of the carriers can be represented in the interband Berry curvature, as well as in the step function

$$\mathbf{J}_\perp(t) = 2\frac{e^2}{\hbar}\mathbf{E}(t) \times \sum_{nm} \int \frac{d^2\mathbf{k}}{4\pi^2} \Omega_{nm}(\mathbf{k} + \Delta k\hat{\mathbf{x}})\Theta(2\omega - \omega_{mn}(\mathbf{k})). \quad (18)$$

As discussed above, this expression is obtained from the first order perturbation expansion of the velocity expectation. Now we consider even further expansion of this first order velocity with respect to the Bloch oscillation amplitude, $\Delta k$. If we assume that the applied field is too weak, so that $\Delta k \to 0$, the Berry curvature can be approximated as follows:

$$\Omega_{nm}(\mathbf{k} + \Delta k\hat{\mathbf{x}}) \approx \Omega_{nm}(\mathbf{k}) + \frac{\partial \Omega_{nm}}{\partial k_x}\Delta k. \quad (19)$$

In the presence of TRS, $\omega_{nm}$ and $\Omega_{nm}$ are even and odd functions of $\mathbf{k}$, respectively, and the integration surface of Eq. (18) contains the pair of time-reversal partner. For each $(\mathbf{k}, -\mathbf{k})$ pair, the summation of Eq. (19) becomes $\Omega_{nm}(\mathbf{k} + \Delta k\hat{\mathbf{x}}) + \Omega_{nm}(-\mathbf{k} + \Delta k\hat{\mathbf{x}}) = 2\frac{\partial \Omega_{nm}}{\partial k_x}\Delta k$. As a result, Eq. (18) can be rewritten as

$$J_y(t) = -\frac{e^3}{\hbar^2\omega}E_0^2 \text{Im}[e^{i2\omega t}]\sum_{nm}\int \frac{d^2\mathbf{k}}{4\pi^2}\frac{\partial \Omega_{nm}}{\partial k_x}\Theta(2\omega - \omega_{mn}), \quad (20)$$

where the multiplication of $\mathbf{E}(t)$ and $\Delta k$ results the frequency doubling. By replacing the integral in Eq. (20) with the interband BCD, we can write the general form as

$$\mathbf{J}_\perp(t) = -\frac{e^3}{\hbar^2\omega}\hat{\mathbf{z}} \times \mathbf{E}(\mathbf{D}_{vc} \cdot \mathbf{E}), \quad (21)$$

which is proportional to Eq. (11) of the main text up to the constant $2\pi$. Hence, we arrive the perturbation theory by weak-field limit of the semi-classical theory.

Now, if we consider the application of strong fields, higher order expansions are needed in Eq. (19). By considering the symmetry arguments above, only odd-order terms survive in the sum of each (**k**, −**k**) pair:

$$\Omega_{nm}(\mathbf{k}+\Delta k\hat{\mathbf{x}})+\Omega_{nm}(-\mathbf{k}+\Delta k\hat{\mathbf{x}}) \approx 2\frac{\partial \Omega_{nm}}{\partial k_x}\Delta k + 2\frac{1}{3!}\frac{\partial^3 \Omega_{nm}}{\partial k_x^3}\Delta k^3 + 2\frac{1}{5!}\frac{\partial^5 \Omega_{nm}}{\partial k_x^5}\Delta k^5 + \ldots, \quad (22)$$

where $\Delta k^3 = -(\frac{eE_0}{2i\hbar\omega})^3(e^{i3\omega t}+3e^{-i\omega t}-c.c.)$ and $\Delta k^5 = -(\frac{eE_0}{2i\hbar\omega})^5(e^{i5\omega t}+5e^{-i3\omega t}+10e^{i\omega t}-c.c.)$.

We can add these contributions into Eq. (22), collect only $2\omega$ SHG terms and can get the higher-order corrected form of Eq. (20) as:

$$J_y(t) = -\text{Im}[e^{i2\omega t}]\left\{\frac{e^3}{\hbar^2\omega}E_0^2(\mathbf{D}_{vc}^{(2)}\cdot\hat{\mathbf{x}}) + \frac{e^5}{12\hbar^4\omega^3}E_0^4(\mathbf{D}_{vc}^{(4)}\cdot\hat{\mathbf{x}}) + \frac{e^7}{384\hbar^6\omega^5}E_0^6(\mathbf{D}_{vc}^{(6)}\cdot\hat{\mathbf{x}}) + \ldots\right\}, \quad (23)$$

A similar form is obtained from the expansion of the 2-band term of the second-order current expectation in the main text.